\documentclass{article}

\begin{document}

\markboth{A.T. Muminov}{Motion of Spin $1/2$ Massive Particle in a
Curved Spacetime}

%
%

\title{Motion of Spin $1/2$ Massive Particle in a
Curved Spacetime}

%

\author{ A.T. Muminov\\
\footnotesize Ulug-Beg Astronomy Institute,
Astronomicheskaya~33,\\\footnotesize Tashkent 100052,
Uzbekistan\\\footnotesize amuminov@astrin.uzsci.net}

\maketitle


\begin{abstract}
Quasi-classical picture of motion of spin $1/2$ massive particle
in a curved spacetime is built on base of simple Lagrangian model.
The one is constructed due to analogy with Lagrangian of massive
vector particle \cite{zr1}. Equations of motion and spin
propagation coincide with Papapetrou equations describing dynamic
of classical spinning particle in a curved spacetime
\cite{papa,zfr}\\{\bf keywords}
Spin $1/2$ massive particle; Dirac
equation; Papapetrou equation.
\end{abstract}

\def \vc {\vec}
\newcommand{\T}{\tilde }
\newcommand{\comment}[1]{}
\newcommand{\lrr}[1]{\left(#1\right)}
\newcommand{\lrc}[1]{\left\{#1\right\}}
\def \wdg {\wedge}
\def \vcx {\dot{\vc x }}
\def \dtx {\dot{x}}
\def \dcn {\mbox{D.C. }}
\newcommand{\lcom}[1]{{\left[#1\right.}}
\newcommand{\rcom}[1]{{\left.#1\right]}}
\newcommand{\lacm}[1]{{\left\{#1\right.}}
\newcommand{\racm}[1]{{\left.#1\right\}}}
\newcommand{\cc}[1]{\bar{#1}}

\newcommand{\ihlf}{\frac{i\hbar}{2}}
\newcommand{\ihqr}{\frac{i\hbar}{4}}
\def \qrt {{1\over4}}
\def \hspin {{1\over2}}
\def \hsp {1/2\,}
\newcommand{\eps}{\varepsilon}
\newcommand{\fracm}[2]
{{\displaystyle#1\over\displaystyle#2\vphantom{{#2}^2}}}
\newcommand{\prt}{\partial\,}
\def \oh {{\tiny 1/2}\,}
\def \moh {{-1/2}}
\def \as {\,^*}
\def \Lge {{\cal L}}
\def \vn {{\vec{n}}}
\def \cdt {\hspace{0.1em}\cdot\hspace{0.1em}}
\def \csdt {\hspace{0.35em}\cdot\hspace{0.35em}}
\def \fdt {\hspace{0.4em}}
\def \tch {\;\grave{}}
\newcommand{\con}[2]{\omega_{#1\cdt}^{\fdt#2}}
\newcommand{\gcn}[2]{\,\grave{}\omega_{#1\cdt}^{\fdt#2}}
\newcommand{\cur}[2]{\Omega_{#1\cdt}^{\fdt#2}}
\newcommand{\tdu}[3]{#1_{\ldots #2\ldots}^{\mdots\ldots #3\ldots}}
\newcommand{\tud}[3]{#1^{\ldots #2\ldots}_{\mdots\ldots #3\ldots}}
\def \gnu {\grave{}\nu}
\newcommand{\dod}[1]{\fracm{\prt}{\prt#1}}
\newcommand{\dd}[2]{\frac{\prt\,#1}{\prt\,#2}}
\newcommand{\pp}[2]{{\prt\,#1}/{\prt#2}}
\newcommand{\dds}[1]{\frac{d\,#1}{ds}\,}
\newcommand{\Ds}[1]{\frac{D\,#1}{ds}\,}
\def \vep {\vc{{}\,\eps}\,}
\def \vn {\vc{{}\,n}}
\def \bps {{\psi^\dagger}}
\def \dps {\dot{\psi}}
\def \dbps {\dot{\bps\,{}}}
\def \dtgam {}
\def \gam {{\hat{\sigma\,{}}\!{}}}
\def \sgm {{\hat{\sigma\,{}}\!{}}}
\def \ggam #1 #2  {\sigma^#1\sigma^#2\,}
\def \onh {$1/2$ }
\def \uno {$1$ }
\def \frm #1  {\{#1\}\,{}}
\def \dgam {}


\section{Introduction}

In terms of classical Lagrange formalism motion of structureless
test particle in curved spacetime is described by simplest form of
Lagrangian:
$$\Lge=1/2\,<\vcx,\vcx>,
$$where $\vcx=\dtx^i\prt_i,$ $\dtx={dx^i\over ds},$ $x^i$ are
coordinates of the particle and $s$ is length along worldline of
the particle. Euler-Lagrange equations for this Lagrangian lead to
the geodesic equation.

In case of non scalar particles additional terms containing spin
variables can be included into the Lagrangian. These internal
variables change equations of motion of the particle due to
spin-gravitational interaction. Frenkel was first who pointed to
fact that spin changes trajectory of motion of particles in
external field \cite{frszk}. Motion of extended spinning particle
in curved spacetime was studied by Papapetrou \cite{papa} and
Dixon \cite{frszk}. Similar problem was studied by Turakulov
\cite{zfr} by means of classic Hamiltonian formalism in
approximation of spinning rigid body in tangent space. In the
mentioned works it was shown that equation of motion differs from
geodesic equation due to term which is contraction the curvature
with spin and velocity. A number of attempts to describe motion of
quantum particles with spin on base of Lagrangian models were made
for last eight decades \cite{frszk}. However, a satisfactory
description was not obtained \cite{zr1}. Nevertheless, studies
both Maxwell and Dirac equations point to the fact that equations
of motion might include contraction spin with the curvature
\cite{d16,d17}. In our recent work \cite{za} a derivation of
Papapetrou equations for photon on base of field variational
principles was completed. In turn, an approach to derive the
equations of motion for spin \uno massive and massless particles
by means of classical Lagrange formalism are shown in our paper
\cite{zr1}.

Particularly, it was shown in the work \cite{zr1} that motion of
massive vector particle of spin $1$ in curved spacetime can be
described by the Lagrangian:
$$
2\Lge=m<\vcx,\vcx>-\dot{\vc{A}}\,\vphantom{\vc{A}}^2+m^2\,{\vc A
}^2,
$$
where $\vc A $ is a vector field which is attached to worldline of
the particle and orthogonal to $\vcx$. Components of the field
expressed in local orthonormal frame are generalized coordinates
which describe spin of the particle.

The goal of present work is to develop a  Lagrangian approach for
spin $1/2$ massive particle. We consider the particle as
quasi-classical. This means that motion of the particle is
described not only by its coordinates $\{x^i\}$ in spacetime but
also by internal spin variables specifying spin degrees of freedom
of the particle in terms of quantum mechanics. It should be noted
that spin variables are elements of suitable spinor spaces. In
turn, determination of the spaces demands presence of orthonormal
frame in considered domain of spacetime. Since particle is massive
we introduce length parameter $s$ along worldline of the particle
which plays role of time parameter in Lagrangian formalism. Thus,
generalized velocities conjugated with coordinates $\{x^i\}$
$\dtx^i=dx^i/ds,$ define timelike vector $\vn_0=\dtx^i\prt_i=\vcx$
of unit length along the worldline:
$$<\vn_0,\vn_0>=1.
$$
There are vectors $\{\vn_\alpha\},$ $\alpha=1,2,3;$ orthogonal to
$\vn_0$ such that
$<\vn_\alpha,\vn_\beta>=\eta_{\alpha\beta}=-\delta_{\alpha\beta}.$
The vectors together with $\vn_0$ constitute comoving frame along
the worldline. Since, by construction, $\vcx$ has no $\vn_\alpha$
components we call spacelike coframe $\{\vn_\alpha\}$ as rest
frame of the particle. Besides, rest frame $\{\vn_\alpha\}$ is
defined with accuracy up to arbitrary spatial rotation belonged to
group $SO(3)$. Due to the fact that generalized coordinates and
velocities of different nature must be independent spinor
variables should be referred to rest frame $\{\vn_\alpha\}$ of
spacelike vectors. In other words, spinor variables are elements
of linear spaces of representation of group $SO(3)$.

The spaces are constructed as follows. Pauli matrices $\frm
\gam^\alpha $ referred to rest frame are introduced. The matrices
generate local Clifford algebras referred to the frame. Besides,
local Clifford algebra introduced this way specifies two local
spinor spaces attached to the worldline. These spaces are two
spaces of spinor representations of the group $SO(3)$ which are
isomorphic to each other under Hermitian conjugation. In our
approach elements of the spaces $\bps,\psi$ play role of
generalized coordinates which describe spin degrees of freedom  of
the particle.

The desired Lagrangian should depend on generalized coordinates
$\frm x^i ,$ $\frm {\bps,\psi} $ and their derivatives over $s$.
At the same time the Lagrangian should contain covariant
derivatives of spinor variables $\bps$ and $\psi$. Moreover, the
Lagrangian must contain term with $<\vcx,\vcx>$ which yields left
hand side (LHS) of geodesic equation. Euler-Lagrange equations for
spinor variables are expected to yield reduced form of Dirac
equation for wave propagating along the worldline of the particle.
In the limiting case of zero gravitation the equation coincides
with Dirac equation formulated in comoving frame of plane spinor
wave of positive energy. All these requirements determine the form
of the Lagrangian describing motion of massive spin \onh particle
in curved spacetime. Thus, Euler-Lagrange equations are reduced to
equations describing motion of the spin along the particle
worldline and the worldline shape. The equations obtained this way
become identical to Papapetrou equations for classic spinning
particle \cite{zfr}.

\section{Lagrange formalism for massive particle of spin $1/2$}

\comment{We decide that motion of spin \onh massive particle like
spin \uno particle can be depicted by means of classical Lagrange
formalism. For this end a set of inner generalized coordinates
describing spin of the particle must be introduces. This aim is
achieved by introduction 1-dimensional spinor fiber bundle on the
worldline of the particle.}
In order to describe spin variables of the Lagrangian we
supplement timelike unit vector $\vn_0=\vcx$ tangent to worldline
of the particle by spatial orthonormal frame $\frm \vn_\alpha $
whose vectors are orthogonal to $\vn_0$. Frame $\frm \vn_\alpha $
is a rest frame of the particle. Spacelike vectors $\vn_\alpha$
together with $\vcx$ constitute orthonormal comoving frame along
the worldline. We denote covector comoving frame as $\frm \nu^a ,$
so $\frm \nu^\alpha$ is covector rest frame dual to vector frame
$\frm \vn_\alpha $ in tangent subspace orthogonal to $\vcx$. Then
we introduce Pauli matrices ${\gam^1,\gam^2,\gam^3}$ referred to
the covector rest frame. The matrices are constant in chosen frame
and obey anticommutation rules as follows:
\begin{equation}\label{gam-rules}
\{\gam^\alpha,\gam^\beta\}=-2\eta^{\alpha\beta},
\end{equation}
where
$$\eta^{ab}=<\nu^a,\nu^b>=diag\,(1,-1,-1,-1)=
\left(
\begin{array}{cc}
1&0\\
0&\eta^{\alpha\beta}
\end{array}
\right).
$$
Algebraic span of Pauli matrices yields local sample of Clifford
algebra in each point of the worldline. Union of the local
Clifford algebras constitute fibre bundle of Clifford algebra
along the worldline.

Invertible elements $R$ of Clifford algebra such that
$$ R^{-1}=R^\dagger,
$$where $R^\dagger$ stands for Hermitian conjugated matrix, constitute
$Spin(3)$ group. There is an endomorphism $R:SO(3)\to Spin(3)$
defined by formula:
\begin{equation}\label{trans}
R_L\gam^aR^{-1}_L=L_{b\cdt}^{\fdt a}\gam^b,\quad (L_{b\cdt}^{\fdt
a})\in SO(3),
\end{equation}
so that each element of $L\in SO(3)$ is covered twice
\cite{ccl,Besse} by elements $\pm R_L\in Spin(3)$.

Elements of local Clifford algebra are operators on two local
spinor spaces referred to considered local frame on the worldline.
The spaces are local linear spaces of representation of group
$Spin(3)$ and $SO(3)$. Elements of the local spaces $\psi\in S$
and $\bps\in S^\dagger$ are $2\times1$ and $1\times2$ complex
matrices accordingly. This way  element $L$ of group of spatial
rotations $SO(3)$ acts on spaces of representation of the group as
follows:
\begin{equation}\label{spstrans} '\psi=R_L\psi,\quad
'\bps=\bps R_L^{-1},\quad \psi\in S,\,\bps\!\in S^\dagger.
\end{equation}
Union of the local spinor spaces constitute spinor fibre bundle on
the worldline.

Image of an infinitesimal rotation $L={\bf1}-\eps\in SO(3)$ is:
\begin{equation}\label{RInf}
R_{1-\eps}=\hat{1}+\delta
Q=\hat{1}+1/4\,\eps_{\alpha\beta}\,\gam^\alpha \gam^\beta.
\end{equation} The infinitesimal transformation rotates elements of
the rest frame:
\begin{equation}\label{rot-cvec}
\delta\nu^\alpha=-\eps_{\beta\cdt}^{\fdt \alpha}\nu^\beta.
\end{equation}
Accordingly (\ref{spstrans}) the rotation initiates a
transformation of spinors:
\begin{equation}\label{spinRot}
\delta\psi=1/8\,\eps_{\beta\gamma}\,[\gam^\beta,\gam^\gamma]\,\psi,\quad
\delta\bps=-1/8\,\eps_{\beta\gamma}\,\bps\,[\gam^\beta,\gam^\gamma],
\end{equation}
under which due to (\ref{trans}) Pauli matrices rotates as
follows:
\begin{eqnarray*}
'\gam^\alpha=R\gam^\alpha R^{-1},\quad \delta\gam^\alpha=[\delta Q,\gam^\alpha];\\
\delta\gam^\alpha=-\eps_{\beta\cdt}^{\fdt \alpha}\gam^\beta=
1/4\eps_{\beta\gamma}\left[\gam^\beta\gam^\gamma,\gam^\alpha\right].
\end{eqnarray*}
It is seen that the rotation coincides with rotation of components
of contravariant vector with accuracy up to opposite sign. Thus,
if we take into account both of the transformations Pauli matrices
become invariant as it is accepted in field theory \cite{Besse}.

State of the particle is described by its coordinates $\frm x^i $
in space time, spinor variables $\frm {\psi,\bps} $ which are
elements of spinor fibre bundles on the worldline and their
derivatives $\dtx^i={dx^i\over ds}$ and $\dds{\psi}$, $\dds{\bps}$
over length $s$ along the worldline. Rest frame rotates under
motion of the particle:
$$\dot\nu^\alpha =-\con{\beta}{\alpha}(\vcx)\nu^\beta ,
$$
where angular velocities are given by values of Cartan' rotation
1-forms $\con{\beta}{\alpha}=\gamma_{c\beta\cdt}^{\fdt\fdt
\alpha}\nu^c$ on vector $\vcx$. So do spinor variables referred to
the frame. Their transformations are given by equations
(\ref{spinRot}) where
$\eps_{\beta\cdt}^{\fdt\alpha}=\gamma_{0\beta\cdt}^{\fdt\fdt\alpha}$.
Account of the transformations are taken by covariant derivatives
of spinor variables along the worldline:
\begin{eqnarray}\label{ders}
\dps=\dds{\psi}+\qrt\gamma_{b\delta\eps}\,\dtx^b\,\gam^\delta\gam^\eps \psi,\\
\dbps=\dds{\bps}-\qrt\gamma_{b\delta\eps}\,\dtx^b\,\bps\,\gam^\delta\gam^\eps.
\nonumber
\end{eqnarray} Besides, total covariant derivatives
(with taking account of spinor transformation and rotation of
vector indexes) of Pauli matrices are zero.

Lagrangian of the particle is covariant under internal
transformations of the rest frames. Hence derivatives (\ref{ders})
are to be included to the Lagrangian. The desired Lagrangian
includes term $m/2\,\bps\psi<\vcx,\vcx>$ which yields geodesic
equation and an addend providing validity of the reduced Dirac
equation. There are also terms including derivatives of spinor
fields and term proportional to $m\bps\psi$. Due to fact that
Dirac equation is of first order partial differential equation the
Lagrangian is to be linear over the derivatives of spinor
variables. Analysis shows that to obey such the requirement we
should accept the form of the Lagrangian:
\begin{equation}\label{lagr}
\Lge={m\over2}\,\bps\psi<\vcx,\vcx>-\ihlf\left(\bps\dps-\dbps\psi\right)
+{m\over2}\,\bps\psi.
\end{equation}
%
%
It must be kept in mind that the Lagrangian is function of
generalized coordinates $x^i,$ $\psi,$ $\bps$ and their velocities
$\dds{x^i}=\dtx^i,$ $\dds{\psi},$ $\dds{\bps}$. At the same time
covariant form of the Lagrangian includes derivatives represented
in orthonormal frame. Due to this we recall formulas of
transformations between the frames:
\begin{eqnarray}\label{frtrans}
\pp{\!}{x^i}=n^a_i\vn_a,\quad\vn_a=n_a^i\pp{\!}{x^i},\quad
n^i_an^b_i=\delta_a^b,\\\nonumber
\dtx^a=n^a_i\dtx^i,\quad\dtx^i=n^i_a\dtx^a,\quad
n^i_an_j^a=\delta^i_j.
\end{eqnarray}

\section{Euler-Lagrange equations for spinor variables}

Due to (\ref{lagr}) generalized momenta  conjugated to generalized
coordinates $\bps$ and $\psi$ are:
$$
\Psi=\pp{\Lge}{\!{}\!\lrr{\dds{\bps}}}=+\ihlf\, \psi,\quad
\Psi^\dagger=\pp{\Lge}{\!{}\!\lrr{\dds{\psi}}}=-\ihlf\,\bps .
$$
Euler Lagrange equations for the considered generalized
coordinates read:
\begin{equation}\label{ELE1}
\dds{}\Psi=\pp{\Lge}{\bps},\quad
\dds{}\Psi^\dagger=\pp{\Lge}{\psi}.
\end{equation}
Straightforward calculation of the right hand side (RHS) of the
above equations gives:
\begin{eqnarray*}
\pp{\Lge}{\bps}=m\psi-\ihlf
\dps-\ihlf\cdot{1\over4}\gamma_{b\delta\eps}
\dtx^b\,\gam^\delta\gam^\eps\psi,\\
\pp{\Lge}{\psi}=m\bps+\ihlf\dbps
-\ihlf\cdot{1\over4}\gamma_{b\delta\eps}
\dtx^b\,\bps\gam^\delta\gam^\eps.
\end{eqnarray*}
Now it is seen that the  RHS of the equations (\ref{ELE1})
completes ordinary derivatives of spinor variables in the LHS up
to covariant derivatives. This way Euler-Lagrange equations for
$\bps,\psi$ generalized coordinates become:
\begin{equation}\label{eq4psi}
i\hbar \dps=m\psi, \quad{}\quad i\hbar\dbps =-m\bps.
\end{equation}
The equations coincide with reduced form of Dirac equations for
free motion of particle with positive energy in flat spacetime
\cite{mss}.

\section{Generalized momentum conjugated with $x^i$ and
conservation of spin}

Due to the definition $p_i=\pp{\Lge}{\dtx^i}.$ However it is
convenient to operate with generalized momenta expressed in
orthonormal frame: $p_a=n_a^ip_i.$ Differentiating (\ref{lagr})
over $\dtx^a$ we obtain:
\begin{eqnarray*}p_a=\dd{\Lge}{\dtx^a}=m\bps\psi
\,\eta_{ab}\dtx^b-\ihlf\cdot{1\over2}\gamma_{a\delta\eps}\,\bps
\gam^\delta\gam^\eps\psi.
\end{eqnarray*}
We define spin of the particle as:
\begin{equation}\label{spin}
S^{\delta\eps}=-\ihqr\,\bps\gam^\lcom{\delta}\gam^\rcom{\eps}\psi,
\end{equation}
where $[,]$ stands for commutator of the Pauli matrices. It is
seen that RHS of the above equation can be represented as
$\hbar/2\,\epsilon_{\delta\eps\zeta}\,\bps\gam^\zeta\psi$, where
$\epsilon_{\delta\eps\zeta}$ is Levi-Civita symbol for the
3-space. In terms of quantum mechanics the expression can be
interpreted as averaged value of operator of spin
$\hbar\gam^\zeta/2$ \cite{mss} in state described by wave function
$\psi$ in the tangent rest space. Moreover, it can be shown that
definition (\ref{spin}) accords to formula for 0-component of
current of spin derived from Noether theorem in field theory
\cite{NN}. The spin is element of space which is tensor product of
two copies of tangent rest space. Thus, it has no 0-component and
we can decide that condition of orthogonality of the spin to
velocity is satisfied: $\dtx^b\,S_{b\cdt}^{\fdt c}\equiv0.$ After
that we can represent expression for the generalized momentum
$p_a$ as follows:
\begin{equation}\label{p_a}
p_a=\pi_a+\hspin\gamma_{a\delta\eps}S^{\delta\eps},
\end{equation}
where $\pi_a=m\,\bps\psi\,\eta_{ab}\dtx^b$ is part of the momentum
including generalized velocities over $x^i$ coordinates.

According to the equation (\ref{eq4psi}) straightforward
calculation of covariant derivative of spin (\ref{spin}) gives:
\begin{equation}\label{DS:ds}
\Ds{S^{\delta\eps}}=\dds{S^{\delta\eps}}
+\dtx^a\lrr{\gamma_{a\gamma\cdt}^{\fdt\fdt \delta}S^{\gamma\eps}
+\gamma_{a\gamma\cdt}^{\fdt\fdt \eps}S^{\delta\gamma}}\equiv0,
\end{equation}
where we took account of the fact that total covariant derivatives
of Pauli matrices are zero.

\section{Euler-Lagrange equations for $x^i$ variables}

Euler-Lagrange equations for $x^i$ variables read:
$$
{dp_i\,/ds}=\pp{\Lge}{x^i}.
$$
It is convenient to rewrite the above equation in orthonormal
frame due to (\ref{frtrans}):
\begin{equation}\label{ELE3}
n^i_a\dds{}\lrr{n^b_ip_b}=n^i_a\,\pp{\Lge}{x^i}.
\end{equation}
After differentiation  and expression velocities $\dtx^k$ via its
components in orthonormal frame the LHS of (\ref{ELE3}) becomes:
\begin{equation}\label{LHS}
n^i_an^b_{i,k}n^k_c\,\dtx^cp_b+\dds{p_a}.
\end{equation}
where $()_{,k}$ means differentiation over $x^k$ variable.
Calculation the RHS of (\ref{ELE3}) gives:
\begin{eqnarray}\nonumber
n^i_a\left[m\bps\psi\,\eta_{bc}n^b_{k,i}n^k_d\,\dtx^d\dtx^c+\ihlf\!\cdot\!{1\over2}
\left(\gamma_{b\delta\eps,i}\,\dtx^b+\gamma_{b\delta\eps}\,
n^b_{k,i}n^k_e\,\dtx^e\right)\,\bps\gam^\delta\gam^\eps\psi=\right.
\\\nonumber
=m\bps\psi\,\eta_{bc}\,n^b_{k,i}n^k_e\,\dtx^e\dtx^c+
\hspin\left(\gamma_{b\delta\eps,i}\,\dtx^b+\gamma_{b\delta\eps}
\,n^b_{k,i}n^k_e\,\dtx^e\right)S^{\delta\eps}=
\\\nonumber
=\left.n^b_{k,i}n^k_e\,\dtx^e(\pi^b+\hspin\gamma_{b\delta\eps}S^{\delta\eps})+
\hspin\gamma_{b\delta\eps,i}\,\dtx^bS^{\delta\eps}\right]=
\\\label{RHS}
=n^b_{k,i}n^i_an^k_e\,\dtx^e\,p_b+\hspin\gamma_{b\delta\eps,a}\,\dtx^bS^{\delta\eps}.
\end{eqnarray}
Equating (\ref{LHS}) to (\ref{RHS}) we obtain:
$$\dds{}p_a+n^b_{i,k}\left(n^i_an^k_c-n^k_an^i_c\right)\dtx^c\,p^b=
\hsp n^i_a\gamma_{b\delta\eps,i}\,\dtx^b\,S^{\delta\eps}.
$$
But due to Cartan' first structure equation it is easy to see
that:
$$n^b_{i,k}\left(n^i_an^k_c-n^k_an^i_c\right)=\gamma_{ac\cdt}^{\fdt\fdt
b}-\gamma_{ca\cdt}^{\fdt\fdt b}.
$$This gives us equation as follows:
\begin{equation}\label{nine}
\Ds{}p_a+\gamma_{ac\cdt}^{\fdt\fdt
b}\,\dtx^cp_b=\hsp\gamma_{b\delta\eps,a}\dtx^bS^{\delta\eps}.
\end{equation}

Since we expect to obtain an equation whose LHS coincides with
geodesic equation generalized momentum $p_a$ in the (\ref{nine})
should be presented in explicit form given by (\ref{p_a}). Leaving
only $D\,\pi_a/ds$  at LHS of (\ref{nine}) we after some simple
derivations can rewrite the equation as:
\begin{eqnarray}\def \cnc #1 #2  {\gamma_{#1\cdt}^{\fdt\fdt#2}}
\nonumber
 \Ds{}\pi_a=-\hspin\dds{}\left[\gamma_{a\delta\eps}S^{\delta\eps}\right]+
\hspin\left[\cnc ba e \gamma_{e\delta\eps}-\cnc ab e
\gamma_{e\delta\eps}+\gamma_{b\delta\eps,a}\right]\dtx^bS^{\delta\eps}=
\\\nonumber
-\hspin\,\gamma_{a\delta\eps}\dds{}S^{\delta\eps}+
\hspin\left[\gamma_{e\delta\eps}d\nu^e+
d\gamma_{e\delta\eps}\wdg\nu^e\right](\vn_a,\vn_b)\,\dtx^bS^{\delta\eps}=
\\\label{Dspi_a}
-\hspin\,\gamma_{a\delta\eps}\dds{}S^{\delta\eps}+
\hspin\,[d\omega_{\delta\eps}](\vn_a,\vn_b)\,\dtx^bS^{\delta\eps}.
\end{eqnarray}
Now equation (\ref{DS:ds}) allows us to exclude
$dS^{\delta\eps}/ds$ from the (\ref{Dspi_a}). After this has been
done the RHS of (\ref{Dspi_a}) turns:
$$\hspin\,\dtx^b\left[\omega_{e\eps}\wdg\con{\delta}{e}(\vn_a,\vn_b)+
d\omega_{\delta\eps}(\vn_a,\vn_b)\right]S^{\delta\eps}=
\hspin\,\dtx^b\Omega_{\delta\eps}(\vn_a,\vn_b)S^{\delta\eps},
$$
where
$$\cur{c}{d}=d\con{c}{d}+\con{e}{d}\wdg\con{c}{e}=1/2R_{c\cdt
ab}^{\fdt d}\,\nu^a\wdg\nu^b
$$ is 2-form of curvature. This way Euler-Lagrange equations for
generalized coordinates $x^i$ become identically with equations:
$$D\pi_a/ds=\hsp R_{\delta\eps ab}\,\dtx^bS^{\delta\eps}.
$$
Substituting $\pi_a=m\bps\psi\,\eta_{ab}\dtx^b$ into the above
equation and reminding that the coefficient at $\eta_{ab}\dtx^b$
is constant we can rewrite the equation as follows:
\begin{equation}\label{papa}
m\bps\psi\,\Ds{\dtx^a}=\hsp R^{\fdt\fdt a}_{\delta\eps\cdt
b}\,\dtx^bS^{\delta\eps}.
\end{equation}
It is seen that (\ref{DS:ds}) and (\ref{papa}) constitute set of
equations of motion of massive particle with spin $s=1/2$ which
coincides with system of Papapetrou equations for motion of
classical spinning particle in curved space-time as they presented
in work \cite{zfr}.

\section*{Acknowledgment}
The author express his gratitude to professor Z. Ya. Turakulov who
motivated the author to carry out this studies and whose critical
remarks provided significant improve of the article. This research
was supported by project FA-F2-F061 of Uzbekistan Academy of
Sciences.


\begin{thebibliography}{99}
\bibitem{frszk}
A.~Frydryszak, {\it Lagrangian Models of Particles with Spin: the
First Seventy Years.} arXive:hep-th/9601020 v.1, 6~Jan (1996)
\bibitem{papa}
Papapetrou A, Proc. R. Soc. {\bf A209,} p248 (1951)
\bibitem{zfr} Turakulov Z Ya, {\it Classical Mechanics
of Spinning Particle in a Curved Space.} arXive:dg-ga/9703008 v.1,
14~March (1997)
\bibitem{d16} P.D. Mannheim, {arXive:gr-qc/9810087.} p21
\bibitem{d17} A.S. Eddington, {\it The Mathematical theory of Relativity,}
Cambridge Univ. Press (1965)
\bibitem{zr1}
Turakulov Z Ya, Safonova M, {\it Motion of a Vector Particle in a
Curved Space-Time. I. Lagrangian Approach.} Mod Phys Lett {\bf
A18} (2003) 579
\bibitem{zr2}
Turakulov Z Ya, Safonova M, {\it Motion of a Vector Particle in a
Curved Space-Time. II. First-Order Correction to a Geodesic in a
Schwarzschild background.} Mod Phys Lett {\bf A~20} (2005) 2785
\bibitem{za}\quad Turakulov Z. Ya, Muminov A. T,
{\it Electromagnetic field with constraints and Papapetrou
equation.} Zeitschrift fur Naturforschung 61{\bf a}, 146 (2006)
\bibitem{ccl} M. Berg, C. DeWitt-Morette, Sh. Gwo and E. Kramer,
{\it The Pin Groups in Physics: C,P and T,} arXive:math-ph/0012006
(2000).
\bibitem{Besse}Seminaire Arthur Besse 1978/79
{\it Geometrie Riemannienne en Dimension 4} CEDIC/FERNAND NATHAN
Paris (1981)
\bibitem{mss} Messiah A. {\it Quantum Mechanics.} {\bf vol.1,2}
New York: J. Wiley \& Sons (1958)
\bibitem{NN}
Bogoliubov N N and Shirkov D V, {\it Introduction to the Theory of
Quantized Fields.} New York: Wiley-Interscience (1959)
\end{thebibliography}
\end{document}